\documentclass[aps,prb,twocolumn,longbibliography]{revtex4-2}
\usepackage{amsmath,amssymb}
\usepackage[pdftex]{hyperref,graphicx}
\hypersetup{colorlinks = true, urlcolor = blue, linkcolor = blue, citecolor = blue}
\usepackage{physics}
\usepackage{xcolor}
\usepackage{bm}
\usepackage{bbold}

\usepackage[normalem]{ulem}

\begin{document}
\title{Random matrix theory for the robustness, quantization, and end-to-end
correlation of zero-bias conductance peaks in a class D ensemble}
\author{Haining Pan}
\affiliation{Condensed Matter Theory Center and Joint Quantum Institute, Department of Physics, University of Maryland, College Park, Maryland 20742, USA}
\author{Jay Deep Sau}
\affiliation{Condensed Matter Theory Center and Joint Quantum Institute, Department of Physics, University of Maryland, College Park, Maryland 20742, USA}
\author{Sankar Das Sarma}
\affiliation{Condensed Matter Theory Center and Joint Quantum Institute, Department of Physics, University of Maryland, College Park, Maryland 20742, USA}

\begin{abstract}
    We develop a general theory to study strong random quenched disorder effects in systems of experimental relevance in the search for Majorana zero modes in topological superconductors.
    Using the random matrix theory in a class D ensemble, we simulate the transport properties of random quantum dots by attaching leads, and calculating the differential conductance in the $S$ matrix formalism. 
    To add the concept of the \textit{length} to the random system so that disordered Majorana nanowires can be simulated by the random matrix theory, we generalize the model of a single quantum dot to a chain of quantum dots by analogy with the superconductor-semiconductor nanowire Majorana platform.
    We first define a new concept, the robustness of zero-bias conductance peaks, in terms of an effective random Hamiltonian considering the self-energy of leads. 
    We then study the joint distribution for the robustness and zero-bias conductance peaks, and find a strong correlation that the zero-bias conductance peak with stronger robustness is also prone to carry a larger conductance peak near $2e^2/h$. 
    This trend is more prominent in shorter chains than in longer chains. This is consistent with experimentally observed zero-bias conductance associated with disorder-induced trivial Andreev bound states (the so-called ugly zero-bias peaks).
    Finally, we study the end-to-end correlation of the disorder-induced zero-bias conductances from two leads by calculating the normalized mutual information, which estimates the degrees of the correlation arising from the trivial zero-bias conductance peaks.  
    Our work provides an estimate of several important metrics used in superconductor-semiconductor experiments to determine the nature of zero-bias conductance peaks, including the robustness, the quantization, and the end-to-end correlation of the trivial zero-bias peaks.
    Therefore, in order to claim any evidence for the Majorana zero modes, one must establish the observed zero-bias conductance peaks to have considerable statistical significance well beyond what we find in this work to exist for the trivial peaks.
\end{abstract}

\maketitle

\section{Introduction}\label{sec:intro}
The experiments using superconductor-semiconductor (SC-SM) nanowires~\cite{sau2010nonabelian,lutchyn2010majorana,sau2010generic,oreg2010helical} searching for the Majorana zero modes (MZMs) have reported many observations of the zero-bias conductance peaks (ZBCPs) since 2012~\cite{deng2012anomalous,mourik2012signatures,das2012zerobias,finck2013anomalous,churchill2013superconductornanowire,lee2014spinresolved,albrecht2016exponential,kammhuber2016conductance,deng2016majorana,zhang2017ballistic,kammhuber2017conductance,chen2017experimental,nichele2017scaling,gul2018ballistic,moor2018electric,vaitiekenas2018effective,zhang2018quantizeda,grivnin2019concomitant,anselmetti2019endtoend,bommer2019spinorbit,chen2019ubiquitous,menard2020conductancematrix,puglia2021closing,yu2021nonmajorana,zhang2021large,song2021large,poschl2022nonlocal,poschl2022nonlocala,wang2022parametric}; in particular, large zero-bias conductance peaks near $2e^2/h$ have been observed recently in Refs.~\onlinecite{nichele2017scaling,zhang2021large,song2021large}.
{Experimental claims have been made repeatedly about ``signatures''~\cite{mourik2012signatures}, ``exponential protection''~\cite{albrecht2016exponential}, ``Majorana bound states''~\cite{deng2016majorana}, ``scaling of peaks''~\cite{nichele2017scaling}, ``quantized conductance''~\cite{zhang2018quantizeda,zhang2021large}, and ``topological superconductivity''~\cite{vaitiekenas2020fluxinduceda} for putative topological MZMs in semiconductor nanowires based on ZBCP observations, which are all, in hindsight, rather obvious disorder effects associated with trivial subgap fermionic states~\cite{bagrets2012class,chiu2017conductance,lai2021theory,liu2017andreev,setiawan2017electron,pan2022ondemand,dassarma2021disorderinduced,pan2020physical,valentini2021nontopological}}. 
However, beyond the abundant observations of ZBCPs in both InSb- and InAs-based nanowire platforms, other hallmarks of the MZMs are yet to be unambiguously confirmed in experiments, including the increasing Majorana oscillation with increasing magnetic field~\cite{dassarma2012splitting}, the bulk gap closing and reopening~\cite{oreg2010helical,sau2010generic,sau2010nonabelian}, the robustness of the quantized ZBCP against gate voltage and magnetic field~\cite{pan2021quantized}, and the end-to-end nonlocal correlation of the ZBCPs from both ends of the wire~\cite{rosdahl2018andreev,lai2019presence,pan2021threeterminal}. 
The absence of these key topological signatures makes just the experimentally observed ZBCPs to be arising from topological MZMs highly unlikely, and many theoretical papers have attributed these ZBCPs to the trivial Andreev bound states arising from the inhomogeneous potential~\cite{kells2012nearzeroenergy,rainis2013realistic,liu2012zerobias,stanescu2014nonlocality,liu2017andreev,reeg2018zeroenergy,degottardi2013majoranaa,moore2018twoterminal,moore2018quantized,pan2020physical,pan2021quantized,stanescu2019robust} or disorder~\cite{bagrets2012class,motrunich2001griffiths,brouwer2011topological,brouwer2011probability,akhmerov2011quantized,lin2012zerobias,neven2013quasiclassical,sau2013density,degottardi2013majorana,roy2013topologically,sau2013bound,stanescu2014nonlocality,adagideli2014effects,hui2015bulk,pan2020physical,pan2021disorder,pan2021crossover,ahn2021estimating} in the system.  While ZBCPs may be necessary for MZMs, they are most certainly not sufficient without compelling evidence for their robustness~\cite{lai2022quality} and their nonlocality~\cite{pan2021threeterminal}.

Although the robustness~\cite{sau2010robustness,brouwer2011topological,lutchyn2011search}, the quantized conductance~\cite{wimmer2011quantum,liu2017andreev}, and the end-to-end correlation~\cite{lai2019presence,rosdahl2018andreev,pan2021threeterminal} beyond topological quantum phase transition (TQPT) are simultaneously manifested by topological ZBCPs, these features may also occasionally, in a limited manner, arise in trivial ZBCPs. Our goal in the current work is to quantify the statistical occurrences of such features induced by trivial ZBCPs using a generic symmetry-based theory assuming the system to be dominated by random disorder. If such quantitative statistical information is available and the experimentally observed ZBCPs manifest the features that are statistically more significant than those revealed by trivial ZBCPs, the likelihood of experimentally observed ZBCPs being the topological MZMs will be greatly enhanced. 
Since the experimental SC-SM nanowire, whether in the topological or trivial phase, is a disordered multichannel system, we expect many of its statistical features to be described by the random matrix theory in a class D ensemble.
Moreover, we are only interested in the most generic statistics of these features, which should depend on the particle-hole symmetry but not on any specific experimental platforms. 
In this work, we, therefore, use the class D random matrix theory to obtain the generic ensemble statistics of trivial ZBCPs in SC-SM Majorana platforms, which should be useful in distinguishing trivial from topological in Majorana experiments in disordered systems.
{The validity of the random matrix theory comes from the number of channels in the system being large, which is thought to be larger than ten. Furthermore, random disorder makes the system decompose into many quantum dots with many channels~\cite{woods2020subband}.}
We have previously worked on the random matrix class D ensemble theory for a single disordered dot, establishing the efficacy of this general approach in understanding experimental ZBCPs~\cite{mi2014xshaped,pan2020generic}.

As a logical continuation of our previous work on the single quantum dot in a class D ensemble~\cite{pan2020generic}, where we answered the question about what kind of apparent quantization of conductances one could expect in a typical nontopological system through postselection and fine-tuning of parameters (extensively employed in making experimental MZM claims), here we study the correlations of the conductances from both ends of the wire, where one has to introduce a pair of leads. 
{The attached leads should be nonidentical as both leads couple to the same set of wavefunctions that are delocalized through the entire dot;}
otherwise, we would have two identical superposed conductance spectra measured from both leads.

In essence, the single quantum dot coupled with two leads represents a very ``short'' Majorana nanowire system; therefore, it will be helpful if we can add the concept of ``{length}'' explicitly to the system by considering a chain of coupled quantum dots. 
We mention that the physical length of a ``short'' wire could actually be long in nanometers or microns because, for small proximity gaps (the current experimental situation), the dimensionless length could still be very short since the SC coherence length (going as the inverse of the gap), which is the unit of length here, could be very long.
In the chain of coupled quantum dots, we attach two leads to both ends of the chain, and assign the concept of the left and right leads to them by analogy with the three-terminal SC-SM nanowire device~\cite{anselmetti2019endtoend,menard2020conductancematrix,puglia2021closing,pan2021threeterminal,poschl2022nonlocal,poschl2022nonlocala,wang2022parametric}. This is a meaningful extension of a quantum dot to a wire as a chain of dots.

Having established the two models that simulate the transport properties of the nanowire experiment, we define a metric called ``\textit{robustness}'' to quantify how robust a ZBCP is against the changes in system parameters. Next, by extracting the joint distribution for the robustness and the zero-bias conductance conditioned on the existence of ZBCPs, we find a strong correlation between them. Namely, the ZBCPs with larger robustness are prone to manifest larger zero-bias conductances near $2e^2/h$, whereas weak robustness is associated with either very small ($\sim 0$) or very large ($\sim 4e^2/h$) zero-bias conductance. This general trend is very prominent in short systems while asymptotically disappears as the system approaches infinity due to the trivial nature of the ZBCPs. We emphasize that most (but not all--- the MZM itself is a very special intrinsic ``perfect'' Andreev bound state occurring precisely at zero energy in a class D topological superconductor and, by contrast, the trivial Andreev bound states are disorder-induced fermionic subgap states which are never precisely at zero energy) of the ZBCPs in this work are, by construction, trivial, arising from disorder-induced subgap fermionic Andreev bound states, and the fact that they so generically resemble the ``expected'' topological MZM signatures for ZBCPs clearly and forcefully reiterates the inherent danger of the experimental work focusing just on fine-tuned ZBCPs as the main evidence for the existence of topological MZMs.  At least, detailed ZBCP statistics must be collected in each experiment before any fine-tuned claims for MZM signatures are made.

Finally, we study the end-to-end correlation of the zero-bias conductances from both leads. Since the conductances from both leads are not guaranteed to be normally distributed, we use the mutual information to quantify the end-to-end correlation instead of the Pearson correlation coefficient which is only meant for normally distributed samples. To compare the mutual information among different samples, we calculate the normalized mutual information (NMI) for each sample by dividing by the corresponding joint entropy. We estimate that the typical NMI is around 0.2 to 0.3 in short systems and approaches zero as the length of the system approaches infinity.
This should also be checked in Majorana experiments by varying the effective wire length, which can be done either by changing the physical length or by changing the applied magnetic field which modifies the induced gap and hence the coherence length.

The rest of the paper is organized as follows. In Sec.~\ref{sec:model}, we construct the theoretical models of the single quantum dot and chain of quantum dots coupled by two leads. We also define several useful metrics and quantities including the {topological visibility} to distinguish the topological phase from the trivial phase, the existence of ZBCPs, and the robustness of ZBCPs, which are fed later into Sec.~\ref{sec:results} to study their statistical distributions.
In Sec.~\ref{sec:results}, we first calculate the robustness and zero-bias conductances as a function of tuning parameters for both the single quantum dot and chain of quantum dots to show their consistent features. We also reveal the correlation between the robustness of ZBCPs and the zero-bias conductance conditioned on the existence of ZBCPs. Finally, we study the end-to-end correlation of the zero-bias conductances from both leads systematically as a function of the system parameters. The discussion and conclusion are presented in Sec.~\ref{sec:conclusion}. In Appendix~\ref{app:A}, we present additional examples of the robustness versus the zero-bias conductance for different models with different parameters. In Appendix~\ref{app:B}, we present additional examples of NMI of zero-bias conductances between both ends for different models with different parameters.

\section{Model}\label{sec:model}
To theoretically simulate the transport properties of the SC-SM nanowire in the presence of strong disorder, we use the random matrix theory in a class D ensemble to describe the Hamiltonian of a random quantum dot. To further simulate the three-terminal devices~\cite{anselmetti2019endtoend,puglia2021closing,menard2020conductancematrix,pan2021threeterminal,poschl2022nonlocal,poschl2022nonlocala,wang2022parametric}, it necessitates two leads attaching to the single quantum dot, or two identical leads attached to both ends of the chain of quantum dots. We first start with the single quantum dot, and review the random matrix theory in a class D ensemble.

\subsection{Single quantum dot}
The Hamiltonian for a single quantum dot in a class D ensemble, governed by the particle-hole symmetry (and no other), is characterized by an $M\times M$ random matrix. For simplicity, the Hamiltonian in the Majorana basis (indexed by m) should satisfy
\begin{equation}
    H_{\text{m}}=-H_{\text{m}}^*.
\end{equation}
Therefore, it is convenient to define $H_{\text{m}}=i A$, where $A$ is a real skew-symmetric matrix. 

In the large-$M$ limit~\cite{mi2014xshaped,guhr1998randommatrix,beenakker1997randommatrix,dittes2000decay}, we assume a Gaussian distribution for $H_{\text{m}}$,
\begin{equation}\label{eq:PH}
    P(H_{\text{m}})\propto \exp(-\frac{\pi^2}{4 M \delta_0} \tr(H_{\text{m}}^2)),
\end{equation}
where $\tr(\cdots)$ takes the trace of the matrix. Here, we set the mean energy-level spacing $\delta_0=1$ for simplicity. 
{This random matrix formalism for a single quantum dot has been discussed before in Refs.~\onlinecite{mi2014xshaped,pan2020generic}.}

\subsection{Chain of quantum dots}
Beyond the single quantum dot, one conceptual novelty in this paper is to explicitly add the ``length'' to the system of quantum dots in order to study the end-to-end correlation of the conductance spectrum similar to the SC-SM nanowire. Therefore, we generalize the single quantum dot to multiple quantum dots which are coupled by the nearest-neighbor hoppings~\cite{iida1990statistical}. In the particle-hole basis, the hopping parameters have opposite signs for the particle channel (red) and the hole channel (green), as shown in Fig.~\ref{fig:2}(a). Thus, we rewrite the Hamiltonian for the single quantum dot in the particle-hole basis (indexed by p-h) through a unitary basis transformation $U$ defined as
\begin{equation}
    H_{\text{p-h}}=U^\dagger H_{\text{m}} U, \quad \text{where}\quad U=\frac{1}{\sqrt{2}} \begin{pmatrix}
        1 & 1 \\
        i & -i
    \end{pmatrix}.
\end{equation}

Therefore, the Hamiltonian for a chain of $L$ quantum dots in the particle-hole basis can be assembled as per
\begin{equation}\label{eq:HL}
    H^L_{\text{p-h}}=\begin{pmatrix}
        H_{\text{p-h}}^{(1)} & T & 0 & \cdots & 0\\
        T & H_{\text{p-h}}^{(2)} & T & \cdots & \vdots\\
        0 & T & H_{\text{p-h}}^{(3)} & \cdots & \vdots\\
        \vdots &\vdots & \vdots &\ddots & T\\
        0 & \cdots & \cdots &T & H_{\text{p-h}}^{(L)}
        \end{pmatrix}.
\end{equation}
Here each $H_{\text{p-h}}^{(i)}$ describing the $i$th quantum dot is a Gaussian random matrix defined by Eq.~\eqref{eq:PH} . $T$ is an $M\times M$ matrix describing the all-to-all couplings that connect every channel within the same particle (hole) sector between the nearest quantum dots, namely,
\begin{equation}
    T=t\begin{pmatrix}
        -\mathbb{1} & \mathbb{0} \\
        \mathbb{0} & \mathbb{1}
    \end{pmatrix},
\end{equation}
where $\mathbb{1}$ is an $M/2\times M/2$ matrix of ones [i.e., $(\mathbb{1})_{ij}=1$], and $\mathbb{0}$ is an $M/2\times M/2$ zero matrix. We also assume $t$ is of the order of the level spacing $\delta_0$. The convention adopted in Eq.~\eqref{eq:HL} is that the first (second) half of indices in $H_{\text{p-h}}^{(i)}$ and $T$ stand for the particles (holes). Thus, for a chain of $L$ quantum dots, the dimension of the Hamiltonian $H^L_{\text{p-h}}$ is $LM\times LM$. 

\subsection{Parametric Hamiltonian}
In real experiments, the SC-SM nanowire is often manipulated by various gate voltages (as well as the applied magnetic field or spin splitting). Therefore, to simulate the fine-tuning process in experiments, we also introduce several tunable ``knobs'' to our theoretical model, where the Hamiltonian is parametrized by $\vec{\alpha}$ as per~\cite{pan2020generic}
\begin{equation}\label{eq:H}
    H= \alpha_1 H_1 + \alpha_2 H_2 +(1-\alpha_1-\alpha_2) H_3,
\end{equation}
where $(H_1,H_2,H_3)$ are three random matrices chosen independently from the Gaussian distribution in Eq.~\eqref{eq:PH}, and the tuning parameter $\vec{\alpha}=(\alpha_1,\alpha_2)$ is a two-dimensional (2D) vector with each component lying between 0 and 1, resembling various gate voltages in experiments. In general, although the number of experimental gates is more than 2, we choose a 2D parameter $\vec{\alpha}$ here only for the purpose of better visualization, and the generalization to a higher dimension of $\vec{\alpha}$ is straightforward [though it is neither necessary nor illuminating since the 2D $\vec{\alpha} $ already suffices to bring out the essential physics, and also the underlying Bogoliubov-de Gennes (BdG) equation for Majorana physics is minimally controlled by two parameters only: chemical potential and Zeeman splitting]. 

Because the Hamiltonian in Eq.~\eqref{eq:H} is parametrized by $\vec{\alpha}$, all physical observables (e.g., the conductance, the robustness of ZBCPs, and the existence of ZBCPs) obtained from the Hamiltonian also depend on $\vec{\alpha}$. Therefore, we present all physical observables in the 2D false-color plot similar to a ``phase diagram'' with the two axes being $\alpha_1 $ and $\alpha_2$. Such a 2D false-color phase diagram is considered one single sample in the total ensemble, where we randomize different sets of $(H_1,H_2,H_3)$ to generate other samples.

\subsection{Leads}
In real experiments, the most common metric measured is the differential conductance $G(V)=dI/dV$ as a function of bias voltage $V$, which indirectly reflects the topological properties of the ZBCPs. In the SC-SM nanowire, the {``quantized''} zero-bias conductance of $2e^2/h$ is the hallmark of the MZMs. However, the reverse is untrue because the {``quantized''} conductance can also be induced trivially by disorder~\cite{pan2020physical,pan2021threeterminal}. Therefore, to simulate the typical metric induced by the random matrix in a class D ensemble, we also calculate the differential conductance in the random matrix formalism by introducing leads to the system. 

{
For a single quantum dot, we attach two leads. Each lead couples $M$ channels in the Hamiltonian to $N$ channels in the lead, as shown in Fig.~\ref{fig:1}(a).  Here, we choose $N=4$ such that the conductance varies between 0 and $4e^2/h$ to simulate the scenario of a single occupied subband in the nanowire in real experiments. Because the distribution of the random matrix $H$ is basis independent, we can choose a particular basis set such that one of the leads $W_1$ is an $M \times N$ diagonal matrix,
\begin{equation}\label{eq:W1}
    (W_1)_{mn}=w_n \delta_{m,n},
\end{equation} 
where the integer index $m \in [1,M]$, integer index $n \in [1,N]$, and $M \gg N$ ($M=80$ in all the following calculations, which are verified numerically to suffice the large-$M$ limit).} The tunneling probability $\Gamma_n \in [0,1]$ controls the coupling strength $w_n$ as per~\cite{guhr1998randommatrix,beenakker1997randommatrix}
\begin{equation}
    \abs{w_n}^2=\frac{M \delta_0}{\pi^2 \Gamma_n} \left(2-\Gamma_n-2\sqrt{1-\Gamma_n}\right).
\end{equation}
In the following calculation, we consider an ideal case by choosing a channel-independent $\Gamma_n=0.1$.
The choice of these parameters does not affect any of our qualitative conclusions.

{
Because the second lead $W_2$ overlaps with the first lead $W_1$, we can construct $W_2$ from the first lead $W_1$ by multiplying an additional orthogonal matrix, i.e.,
\begin{equation}\label{eq:W2}
    W_2=O W_1,
\end{equation}
where $O$ is an $M\times M$ orthogonal matrix chosen in a way that $W_2$ does not commute with $W_1$, i.e., $\left[W_2 W_2^\dagger,W_1 W_1^\dagger\right]\neq 0$. Numerically, it makes no difference to choose any random orthogonal $O$. However, once we choose an orthogonal matrix $O$, we fix it such that $W_1$ and $W_2$ are also invariant while tuning $\vec{\alpha}$ in Eq.~\eqref{eq:H} under a fixed set of $(H_1,H_2,H_3)$. For different sets of $(H_1,H_2,H_3)$, we randomize $W_2$ again by choosing a new random orthogonal matrix $O$.
}

For the chain of quantum dots, we attach two leads to the two quantum dots lying on both ends of the chain, as shown in Fig.~\ref{fig:2}(a). Again, it is easy to choose the basis of the first quantum dot [QD$_1$ in Fig.~\ref{fig:2}(a)] and the last quantum dot [QD$_L$ in Fig.~\ref{fig:2}(a)] individually such that both $W_L$ and $W_R$ are diagonal matrices. We also set the identical tunneling probability $\Gamma_n$ for both leads, which allows us to simply define $W_{R}\equiv W_{L}$ according to Eq.~\eqref{eq:W1}. 

\subsection{Differential conductance}
To calculate the differential conductance measured from both leads, we need to evaluate the $S$ matrix by applying the Mahaux-Weidenm\"uller formula~\cite{guhr1998randommatrix,beenakker1997randommatrix,mahaux1969shellmodel,christiansen2009mathematical,marciani2014timedelay}, where we conceptually treat two leads as one whole lead denoted by $W$, i.e.,

\begin{equation}
    W_{M\times 2N}=\begin{pmatrix}
        W_1 & W_2
    \end{pmatrix} 
\end{equation} for the single quantum dot, 
and
\begin{equation}
    W_{LM\times 2N}=\begin{pmatrix}
        W_L & 0\\
        \vdots & \vdots\\
        0 & W_R
    \end{pmatrix}
\end{equation}
for the chain of quantum dots.

Thus, the unitary $S$ matrix at an energy $E$ is
\begin{equation}\label{eq:mw}
    S(E)=\mathbf{1}_{2N}+2i\pi W^\dagger \left(H- i\pi W W^\dagger -E\right)^{-1} W,
\end{equation}
where $\mathbf{1}_{2N}$ is a $2N \times 2N$ identity matrix.

Therefore, the conductance of a random Hamiltonian in the particle-hole basis at a bias voltage $V$ from Lead$_i$ is given by
\begin{equation}
    G_{i}(V)=\frac{e^2}{h}\left[\frac{N}{2}-\frac{1}{2} \tr([S(eV)]_{ii}\tau_z [S^\dagger(eV)]_{ii} \tau_z)\right],
\end{equation}
where $i=\{1,2\}$ for the single quantum dot, $i=\{\text{L},\text{R}\}$ for chain of quantum dots, and $e$ is the electron charge. $[S(eV)]_{ii}$ denotes the reflection matrix on the diagonal block of the $S$ matrix, and the Pauli matrix $\tau_z$ acts on the particle-hole space.

\subsection{{Topological visibility}}
Although the differential conductance is the only measurable quantity in tunneling spectroscopy in experiments, indirectly reflecting the Majorana topology, we have an advantage in doing the theory that we can directly determine the topology. {Here by ``topology'' we mean a system with the nontrivial coupling to leads in $S$ matrix formalism~\cite{pikulin2013two}.}
Since we have already calculated the $S$ matrix, it is straightforward to define the topological visibility $Q_i$ {based on the determinant} of the reflection matrix at zero energy from Lead$_i$, i.e.,
{\begin{equation}
    Q_i=\det([S(E=0)]_{ii}),
\end{equation}}
where $Q_i=+1$ ($Q_i=-1$) indicates the trivial (topological) phase {in the ideal infinite system}~\cite{akhmerov2011quantized,dassarma2016how}. In general, $Q>0$ is always trivial, and $Q \sim -1$ is topological. 
{However, any finite system such as the one considered in this work is technically zero dimensional and therefore nontopological by definition. However, for practical use, it is necessary to classify finite systems as topological, since all systems in nature are finite. Thus, the scattering matrix topological invariant~\cite{akhmerov2011quantized,fulga2011scattering,fulga2012scattering} is computed for finite lattice systems using leads attached to both ends. The finite system is then typically~\cite{pekerten2017disorderinduced,choy2011majorana,cole2017ising} declared as topological for values $Q=-1$ of the scattering matrix invariant. As discussed in the Appendix of Ref.~\onlinecite{akhmerov2011quantized}, the assignment $Q=-1$ in this system represents the case where the coupling of the leads to the system is stronger than the splitting between the end Majorana modes. We follow this standard convention for the definition of a finite system as topological that is used in the literature. While the main goal of our work is to establish the possibility of quantized conductance in a nontopological phase, it is important to note that the quantum dot chain Hamiltonian considered in this work is known to be topological in certain limits~\cite{sau2012realizing,fulga2013adaptive,zhang2016majorana}.  The conductance, in this case, is also quantized. Since we are interested in understanding when conductance is quantized in the putative nontopological case, we compute $Q$ to separate the cases $Q=1$ and $Q=-1$.}
{Therefore, in practice, we adopt a cutoff of $Q<-0.95$ as a criterion to indicate the topological phase.} In the following results, we will explicitly show {by numerics} that it does not matter from which end we calculate the {topological visibility} as they are invariably identical. Therefore, we simply use $Q$ to denote the {topological visibility} from both leads.

\subsection{Distribution of zero-bias conductance peaks}
Besides the differential conductance, we are also interested in the distribution of ZBCPs. Therefore, we introduce a non-Hermitian effective Hamiltonian from Eq.~\eqref{eq:mw}, 
\begin{equation}\label{eq:Heff}
    H_{\text{eff}}=H-i \pi W W^\dagger,
\end{equation}
where the imaginary part describes the self-energy of the leads. We note that all eigenvalues of $H_{\text{eff}}$ lie in the lower half of the complex plane due to the positive definiteness of $W W^\dagger$. In addition, the particle-hole symmetry implies that for any eigenvalue $\epsilon $ of $H_{\text{eff}}$, $-\epsilon^*$ must appear as another eigenvalue as well. Such a constraint results in the fact that all eigenvalues are symmetrically distributed around the imaginary axis in the complex plane unless they are exactly on the imaginary axis. This symmetric distribution of eigenvalues of $H_{\text{eff}}$ is the origin of the robustness of ZBCPs in a class D ensemble against the small perturbation because the purely imaginary eigenvalue cannot acquire a finite real component without breaking the particle-hole symmetry~\cite{pikulin2012topological}. Thus, by identifying the purely imaginary eigenvalues of the non-Hermitian $H_{\text{eff}}$ in Eq.~\eqref{eq:Heff} as a function of $\alpha_1$ and $\alpha_2$, we can determine whether a ZBCP exists and, thus, obtain a ``phase diagram'' for ZBCPs (e.g., the red contours in Figs.~\ref{fig:1} and~\ref{fig:2}).

\subsection{Robustness}
Since the robustness of zero-energy modes arises from the particle-hole symmetry in a class D ensemble, an interesting question is whether this robustness of the eigenvalues of the effective Hamiltonian Eq.~\eqref{eq:Heff} (equivalently, the robustness of the ZBCP) can be directly quantified. It is important because this may provide a quantitative estimation of the robustness of the trivial ZBCPs. Because the real MZMs exhibit stronger robustness against system parameters such as the Zeeman field, tunnel gate, etc, the experimental observations should manifest stronger robustness than that created by trivial ZBCPs in a class D ensemble in order to claim the existence of topological MZMs. This issue of the true robustness of the observed ZBCPs was the key physics involved in the retraction of a recent experimental claim for Majorana quantization~\cite{zhang2018quantizeda}.  We believe that all existing ZBCP-based experimental MZM claims, bar none, suffer from the lack of requisite topological stability, and are most likely class D trivial peaks arising from disorder in the system~\cite{pan2020physical,pan2021quantized,ahn2021estimating,woods2021chargeimpurity}.

Therefore, we define the robustness $R$ of ZBCPs in the 2D ``phase diagram'' [e.g., Fig.~\ref{fig:1}(b)] as
\begin{equation}\label{eq:R}
    R(\vec{\alpha})=\begin{cases}
        \min\limits_{\vec{\alpha}'\in \partial B} \left\| \vec{\alpha}-\vec{\alpha}' \right\| & \text{for} \quad \vec{\alpha} \in B\\
        0 & \text{for} \quad \vec{\alpha} \notin B.
    \end{cases}
\end{equation}
Here, $B$ is the set where ZBCPs exist, and $\partial B$ is its boundary, and $\left\| \vec{\alpha}-\vec{\alpha}' \right\|$ measures the Euclidean distance between $\vec{\alpha}$ and $\vec{\alpha}'$ in the parameter space. For each point at $\vec{\alpha}$, the robustness simply measures the shortest distance from this point (inside the set of ZBCPs) to the boundary of the set of ZBCPs. For the points in the parameter space that do not have ZBCPs, we just assign the robustness to be zero. An example of the phase diagram of robustness measure is shown in Fig.~\ref{fig:1}(b). 

Because the zero-bias conductance is also a function of $\vec{\alpha}$, another interesting question is the statistical correlation between the robustness $R(\vec{\alpha})$ and the zero-bias conductance $G(V=0;\vec{\alpha})$, namely, whether the trivial ZBCPs with large conductances are prone to manifest more robustness and vice versa. Therefore, by estimating the joint distribution for $R$ and $G$ conditioned on the existence of ZBCPs as well as the conditional distribution $P(G|R=R_0)$, where $R_0>0$, we can directly visualize their correlations. Similarly, by analyzing the experimental observations to extract the same metric of the robustness, and comparing them with the statistics obtained from trivial ZBCPs, we may gain further insights into the nature of experimentally observed ZBCPs.
\subsection{End-to-end correlation}

Now, we consider the end-to-end correlation between the two leads by studying their conductances in the single quantum dot as well as in the chain of quantum dots. To quantify the end-to-end correlation of the zero-bias conductances between the two leads, we adopt the mutual information instead of the Pearson correlation coefficient because the latter one is meant for normally distributed samples only, which the conductances do not always satisfy without \textit{a priori} knowledge. Therefore, we resort to a correlation measure that does not rely on any \textit{a priori} knowledge of the underlying distribution, namely, the mutual information that quantifies how much information we can infer about the second lead if we are only given the conductances measured from the first lead. The mutual information of the two sets of zero-bias conductances from both leads is defined as
\begin{equation}
    I(\{G_i\};\{G_j\})=S(\{G_i\})+S(\{G_j\})-S(\{G_i\} , \{G_j\}),
\end{equation}
where $\left( i,j \right)=\left( 1,2 \right)$ for the single quantum dot, and $\left( i,j \right)=\left( L,R \right)$ for the chain of quantum dots. $S(\{G_{i}\})$ is the entropy of $\{G_i\}=\{G_{i}(V=0;\vec{\alpha}) | \vec{\alpha}\in \left[0,1\right]\times \left[0,1\right]\}$ and $S(\{G_i\} , \{G_j\})$ is the joint entropy of $\{G_{i}\}$ and $\{G_{j}\}$.

Furthermore, in order to compare the mutual information among different samples, we normalize the mutual information of each sample by dividing by the joint entropy of that sample, namely,
\begin{equation}
    \text{NMI}(\{G_i\};\{G_j\})=I(\{G_i\};\{G_j\})/S(\{G_i\} , \{G_j\}).
\end{equation} 
Thus, the NMI is unity if both leads measure the same conductances, and zero if they measure completely uncorrelated conductances.

\section{Results}\label{sec:results}
In this section, we present our numerical results generated from a large ensemble [1000 different sets of $(H_1,H_2,H_3)$]. We start with the single quantum dot with two leads as shown in Fig.~\ref{fig:1}, and generalize it to the chain of quantum dots in Fig.~\ref{fig:2}.

\subsection{Single quantum dot}

\begin{figure}[ht]
    \centering
    \includegraphics[width=3.4in]{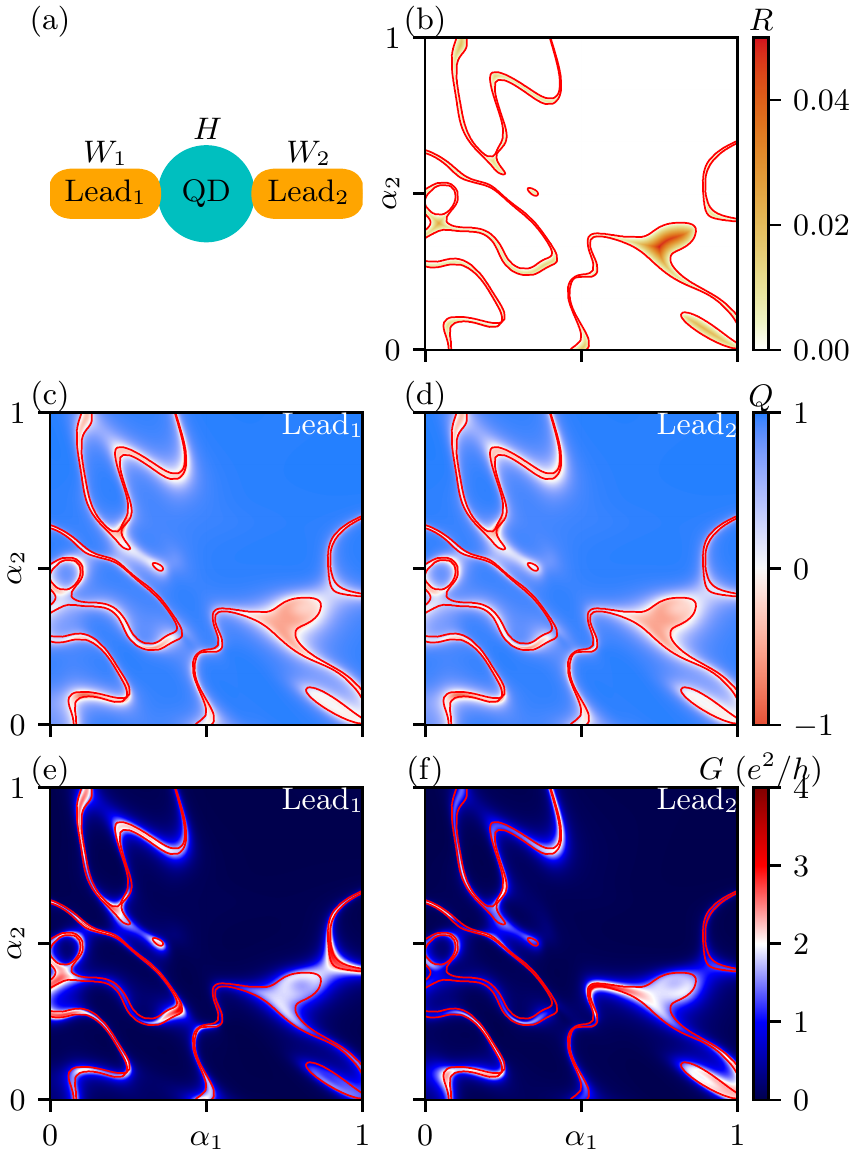}
    \caption{(a) The schematic of a single quantum dot $H$ in a class D ensemble coupled by two leads $W_1$ and $W_2$. 
    (b) The robustness of ZBCPs as a function of $\alpha_1$ and $\alpha_2$. The red contours [also in (c-f)] represent the boundary of the set of ZBCPs with reddish regions being more robust ZBCPs. 
    (c-d) {topological visibility} $Q$ as a function of $\alpha_1$ and $\alpha_2$ measured from Lead$_1$ and Lead$_2$.
    (e-f) Zero-bias conductances $G$ as a function of $\alpha_1$ and $\alpha_2$ measured from Lead$_1$ and Lead$_2$. 
    }
    \label{fig:1}
\end{figure}

In Fig.~\ref{fig:1}(a), we present the schematic plot where $H$ is the Hamiltonian for the random matrix in a class D ensemble with $W_1$ and $W_2$ being the coupling matrices of two leads. We fix a particular set of $(H_1,H_2,H_3)$ while tuning $(\alpha_1,\alpha_2)$ to generate a 2D ``phase diagram''.

In calculating the robustness $R$ as a function of $\alpha_1$ and $\alpha_2$ as shown in Fig.~\ref{fig:1}(b), we first solve the effective Hamiltonian in Eq.~\eqref{eq:Heff} to search for the existence of ZBCPs by identifying the purely imaginary eigenvalues of $H_{\text{eff}}$, where their boundaries are labeled in red contours. Inside the region where ZBCPs exist, we then calculate the shortest center-to-boundary distance of each point based on Eq.~\eqref{eq:R}, where red indicates stronger robustness. Figure~\ref{fig:1}(b) shows that most of the ZBCPs exist in very narrow ridgelike regions~\cite{pan2020generic} and therefore are not robust against small perturbations of $\alpha_1$ and $\alpha_2$.  

However, ZBCPs can occasionally exist in a large plateaulike region collectively, e.g., near the center of Fig.~\ref{fig:1}(b) at $\vec{\alpha}=(0.75,0.4)$. If the parameters happen to fall inside the plateaulike region, by fine-tuning parameters $\alpha_1$ and $\alpha_2$ to the center of that plateaulike region, we can observe ZBCPs with a considerable degree of robustness ($R\sim 0.05$) that are significantly stronger than the other unstable trivial ZBCPs ($R<0.01$). However, from the {topological visibility} in Figs.~\ref{fig:1}(c) and~\ref{fig:1}(d), we notice that $Q$ is still far from the $-1$, which means that they are still trivial ZBCPs although occasionally these trivial zero modes may form a misleading somewhat-stable ``phase''. In real experiments, due to their relatively strong robustness against parameters, these trivial ZBCPs with $R\sim0.05$ may resemble topological ZBCPs if they also happen to show the zero-bias conductances near the {``quantized''} value of $2e^2/h$.

Therefore, we present the corresponding zero-bias conductances in Figs.~\ref{fig:1}(e) and~\ref{fig:1}(f) from Lead$_1$ and Lead$_2$, respectively. We find that zero-bias conductances near $\vec{\alpha}=(0.75,0.4)$ also happen to be nearly {``quantized''} at $2e^2/h$, which exemplifies the false-positive character of topological ZBCPs manifested by trivial ZBCPs.  
In Figs.~\ref{fig:1}(e) and~\ref{fig:1}(f), we also indicate the boundaries of trivial ZBCPs using red contours. Therefore, combining with the 2D trivial ``phase diagram'' showing the zero-bias conductances, we notice that if the parameter $\vec{\alpha}$ is outside the set of ZBCPs, the absence of a zero-energy mode generally results in a gapped phase with a vanishing zero-bias conductance. However, if $\vec{\alpha}$ is inside the set of ZBCPs, the corresponding zero-bias conductance can, in principle, be any value between 0 and $4e^2/h$~\cite{pan2022ondemand}, and in particular ``unlucky'' situations may just happen to be $2e^2/h$  in spite of being totally trivial.  This is bad news because this means that fine-tuning and postselection, used extensively in the experiments, could lead to misleading conclusions when focusing just on ZBCPs and their values, as is now already well established through direct realistic microscopic calculations~\cite{pan2020physical}.

More specifically, for ZBCPs in the ridgelike region with very weak robustness $R$, we find that their conductances are also relatively small [$G(V=0)\ll e^2/h$]. This feature can be understood as these peaks are very unstable--- they will disappear and become gapped states even if the parameters slightly change, and the zero-bias conductance of a gapped state is exponentially small. Thus, there is no way for the conductance to change drastically from a vanishing value induced by a gapped phase to a large conductance arising from a zero-energy state. 
However, the stable trivial ZBCPs [e.g., the center of Fig.~\ref{fig:1}(b) near $\vec{\alpha}=(0.75,0.4)$] show an opposite trend: they can manifest relatively large zero-bias conductances. Near the center of the plateaulike region with $R\sim 0.05$, we find the zero-bias conductances happen to be nearly {``quantized''} at $2e^2/h$. This connection between the robustness $R$ and zero-bias conductance $G$ of the trivial ZBCPs is intriguing, and worth a systematic study, as we show in later sections. 

\subsection{Chain of quantum dots}

\begin{figure}[ht]
    \centering
    \includegraphics[width=3.4in]{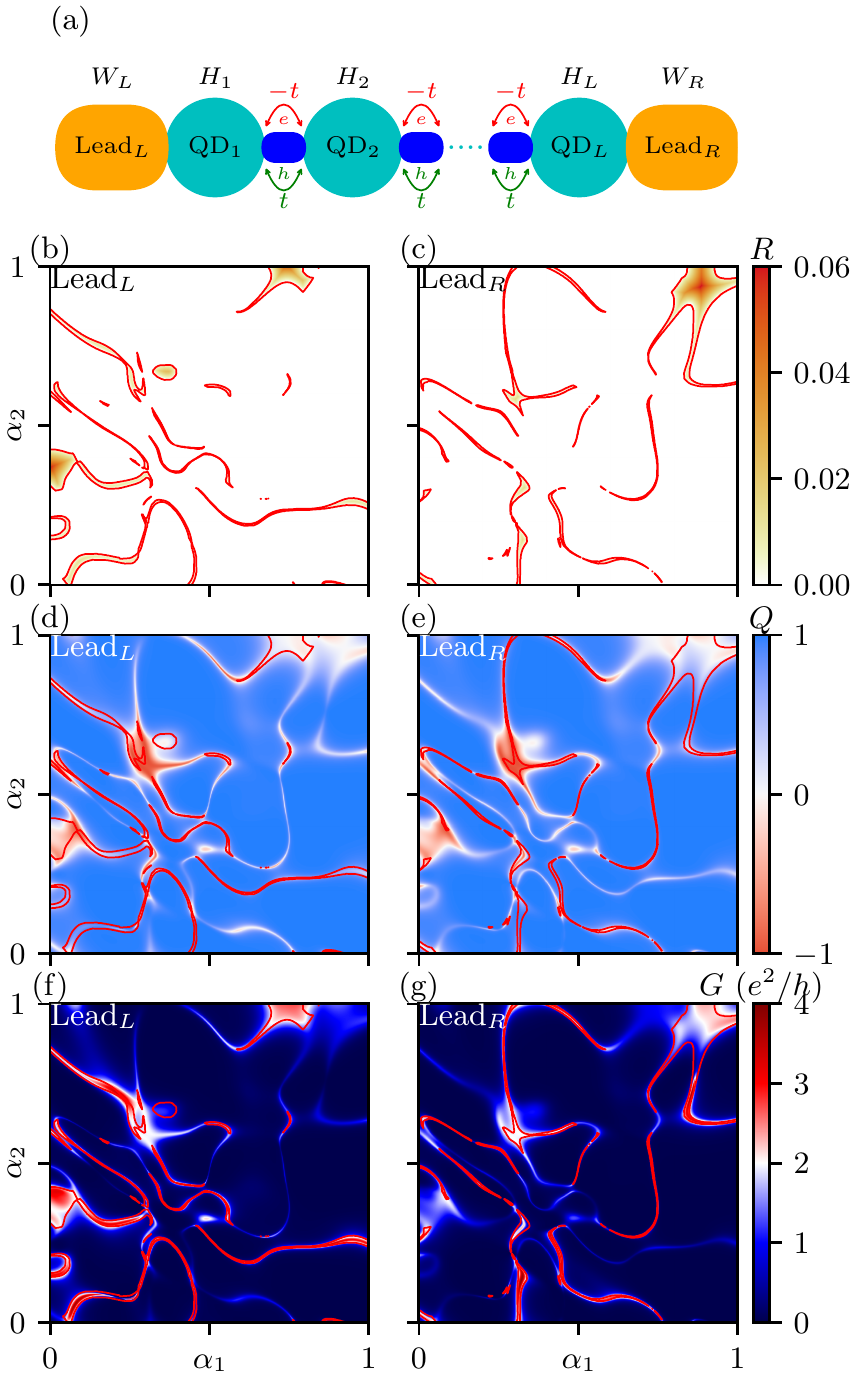}
    \caption{(a) The schematic of a chain of quantum dots in a class D ensemble coupled by two identical leads on opposite ends of the chain. The hopping between nearest neighbor quantum dots is $-t$ ($t$) for the electron (hole) channel.
    (b-c) The robustness of ZBCPs as a function of $\alpha_1$ and $\alpha_2$ from the left and right leads. The red contours [also in (d-g)] represent the boundary of the set of ZBCPs that are more localized at the left end (shown in the left column) or the right end (shown in the right column). 
    (d-e) {topological visibility} $Q$ as a function of $\alpha_1$ and $\alpha_2$ measured from Lead$_L$ and Lead$_R$.  
    (f-g) Zero-bias conductances as a function of $\alpha_1$ and $\alpha_2$ measured from Lead$_L$ and Lead$_R$.
    }
    \label{fig:2}
\end{figure}

Before delving into the connection between the robustness $R$ and zero-bias conductance $G$, here we present another example, the chain of quantum dots, to show that these ZBCP features are generic in a class D ensemble, and do not depend on the specific model. Therefore, we construct a chain of $L$ quantum dots, where each pair is coupled by the identical nearest hopping, as shown in Fig.~\ref{fig:2}(a). At two ends of the chain, we attach two leads to measure the conductances. Here we choose the simplest case by setting $L=2$ and plot the ``phase diagrams'' of the robustness [Figs.~\ref{fig:2}(b) and~\ref{fig:2}(c)], {topological visibility} [Figs.~\ref{fig:2}(d) and~\ref{fig:2}(e)], and zero-bias conductances [Figs.~\ref{fig:2}(f) and~\ref{fig:2}(g)] as a function of $\alpha_1$ and $\alpha_2$.

To calculate the robustness $R$ [Figs.~\ref{fig:2}(b) and~\ref{fig:2}(c)] similarly in the chain of quantum dots, we also need to define the set of ZBCPs first. However, unlike the previous single quantum dot, where the zero-energy state can be probed by both ends simultaneously, here we need to specify the localization of a ZBCP, i.e., at which lead the peak is more likely to be probed. 
Therefore, we compare the zero-bias conductances measured from the left and right leads, and choose the end with the larger zero-bias conductance to determine the localization of a ZBCP. 
Because a larger conductance at a specific end indicates that the zero-energy state mostly spatially resides at that end due to the positive correlation between the local conductance and the local density of states, we can effectively assign the ZBCP to that particular end. 
By that, we decompose the entire ``phase diagram'' of ZBCPs, which is extracted by identifying purely imaginary eigenvalues of $H_{\text{eff}}$ into two ``phase diagrams'' of ZBCPs from the left and right ends, respectively, as shown in Figs.~\ref{fig:2}(b) and~\ref{fig:2}(c).

We find similar behaviors of the robustness in these ``phase diagrams'' to those in the single quantum dot: most of the ZBCPs in the chain of quantum dots lack robustness against the changes of parameters; occasionally, a plateaulike region can appear and manifest robust trivial ZBCPs [e.g., $R\sim 0.06$ near the upper right of Figs.~\ref{fig:2}(b) and~\ref{fig:2}(c) at $\vec{\alpha}=(0.9,0.9)$]. However, one distinction in the chain of quantum dots, in contrast to our single-dot results where almost all ZBCPs are trivial, is the possibility of a finite region where the topological phases exist, e.g., near $\vec{\alpha}= (0.3,0.6)$ as shown in Figs.~\ref{fig:2}(d) and~\ref{fig:2}(e).

In Figs.~\ref{fig:2}(f) and~\ref{fig:2}(g), we also calculate ``phase diagrams'' of the zero-bias conductances from both ends. First, for topological phases, zero-bias conductances are invariably {``quantized''} at $2e^2/h$, which is the typical hallmark of Majorana zero modes (at $T=0$). 
Second, for trivial phases, we compare the ``phase diagram'' of zero-bias conductances with that of the robustness at the same end and reach a similar conclusion as before that the unstable trivial ZBCPs statistically manifest small zero-bias conductances, while the stable trivial ZBCPs show much larger zero-bias conductances near $2e^2/h$. This common feature in both the single quantum dot and chain of quantum dots strongly indicates the ubiquity of the correlation between the robustness and zero-bias conductances, which we will discuss in the next section.

\subsection{Correlation between $R$ and $G$}

\begin{figure}[ht]
    \centering
    \includegraphics[width=3.4in]{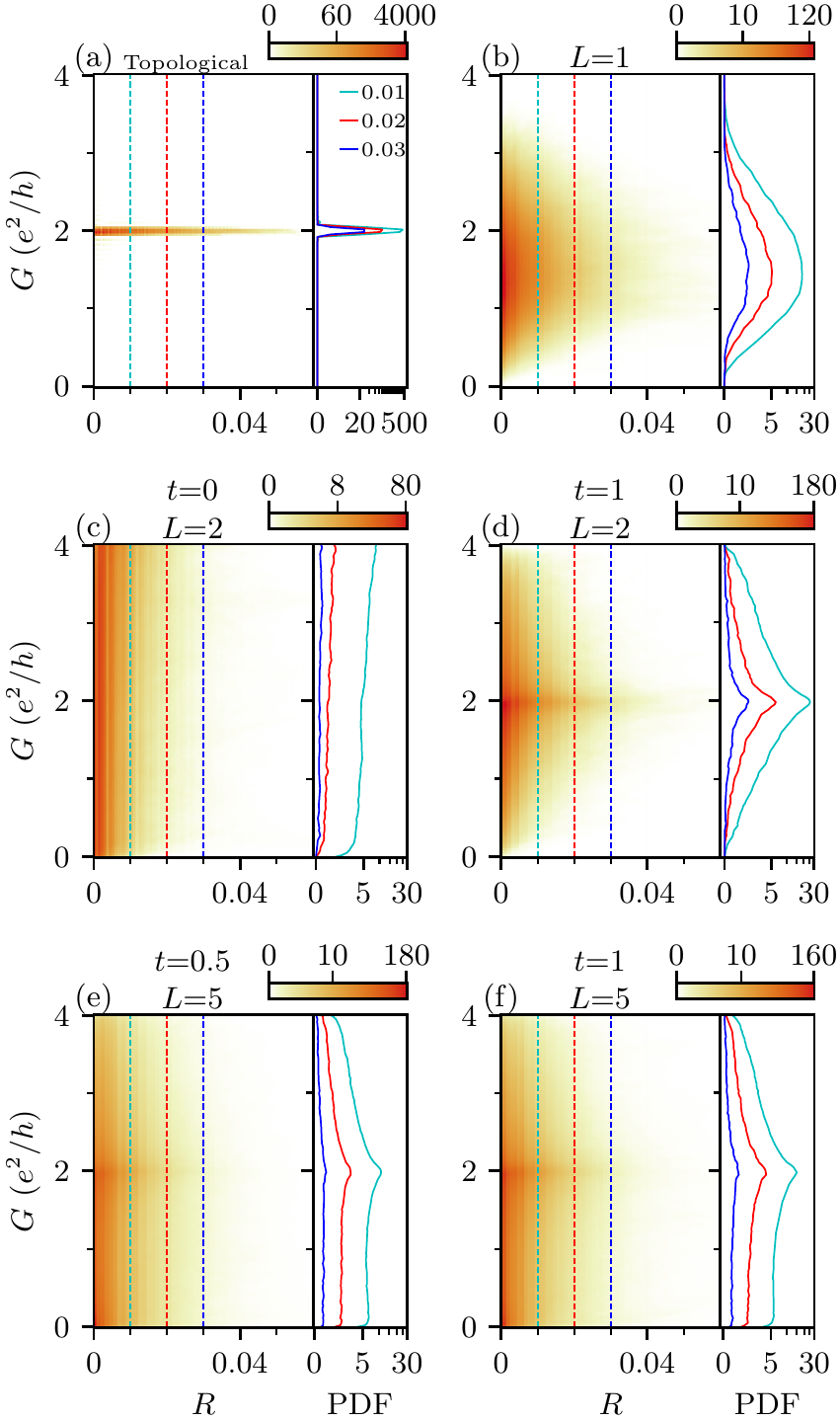}
    \caption{The joint distribution for the robustness $R$ and the zero-bias conductance $G$ for (a) a topological case (as a reference), (b) a single quantum dot with two leads; (c) two quantum dots with no coupling, (d) two quantum dots with $t=1$; (e) five quantum dots with $t=0.5$, (d) five quantum dots with $t=1$. The solid lines on the right-hand sides of each panel show the conditional probability density function at a fixed robustness $R=0.01$ (cyan), $R=0.02$ (red), and $R=0.03$ (blue). The probability density functions are all rescaled logarithmically for better visualization. The ensemble size for each case is 1000.}
    \label{fig:3}
\end{figure}

In both the single quantum dot (Fig.~\ref{fig:1}) and the chain of quantum dots (Fig.~\ref{fig:2}), we seem to find a correlation between the robustness and zero-bias conductances in the trivial ZBCPs: the weak robustness is correlated to the vanishing zero-bias conductance, while the strong robustness is correlated to a large zero-bias conductance near $2e^2/h$. However, very large (i.e., the maximally allowed) zero-bias conductances ($G~\sim 4e^2/h$) are also rare. Therefore, to fully understand the correlation quantitatively, we take each pair of $R(\vec{\alpha})$ and $G(\vec{\alpha})$ at the same $\vec{\alpha}$ from both the ``phase diagrams'' of the robustness and the ``phase diagrams'' of the zero-bias conductances, filter out the points of $\vec{\alpha}$ that can host ZBCPs (i.e., $R>0$), and estimate the probability density function (PDF) of the joint distribution for $R$ and $G$ conditioned on the existence of ZBCPs [i.e., $P(G,R | R>0)$] using the multivariate kernel density estimate (KDE) with the Gaussian kernel, along with the corresponding conditional distribution $P(G|R=R_0)$ given fixed robustness $R_0$. In Fig.~\ref{fig:3}, each panel represents a unique ensemble from different models (the single quantum dot or the chain of quantum dots) or different parameters ($t$ and $L$ in the chain of quantum dots). 

We first present an example of a topological case as a reference in Fig.~\ref{fig:3}(a), which is {a collection of all topological points} (i.e., $\left\{\vec{\alpha}|Q(\vec{\alpha})<-0.95\right\}$, where $-0.95$ a numerical cutoff instead of $-1$) in the ensemble from $L=2$ and $ t=1$ (there is nothing special about this ensemble as all topological phases result in a similar joint distribution). In the left-hand side of Fig.~\ref{fig:3}(a), we present the PDF of the joint distribution for $R$ and $G$ with the false color rescaled logarithmically (i.e., $P\mapsto\log(P+1)$, and +1 is simply added to avoid divergence at zero probability) to better visualize the peak. We note that all zero-bias conductances have a sharp localization at the {``quantized''} $2e^2/h$, and their robustness can generally extend to as large as $R\sim 0.06$. Here the robustness is still defined based on Eq.~\eqref{eq:R}. In the right-hand side of Fig.~\ref{fig:3}(a), we show the three conditional distributions of $P(G|R=0.01)$ (cyan), $P(G|R=0.02)$ (red), and $P(G|R=0.03)$ (blue), which all manifest Lorentzian distributions centered at $2e^2/h$.

Second, we show the PDF of the joint distribution for $R$ and $G$ in the single quantum dot with two leads in Fig.~\ref{fig:3}(b). We find that the joint distribution for $R$ and $G$ manifests a dome shape along the $G$ axis, where the large conductance ($G>3e^2/h$ ) and small conductances ($G<e^2/h$) are both statistically rare, and most samples peak at the conductances near $1.5\sim 2 e^2/h$. In terms of the robustness $R$, it is very weak at large and small conductances, which is consistent with our previous observations in Figs.~\ref{fig:1} and~\ref{fig:2}. On the other hand, most of the samples with a zero-bias conductance between $1.5e^2/h$ and $2e^2/h$ manifest much stronger robustness, in spite of being trivial, up to $R\sim 0.03$. 

Beyond the single quantum dot with two leads, we also study the chain of quantum dots. As a reference, we first present the PDF of the joint distribution for $R$ and $G$ in two decoupled quantum dots with $t=0$. Unlike the topological case in Fig.~\ref{fig:3}(a), the two decoupled quantum dots are effectively two copies of a single quantum dot each with only \textit{one} lead attached, {which is always manifestly trivial according to the unitarity of the scattering matrix topological invariant}~\cite{pan2020generic}. We find that zero-bias conductances conditioned on the robustness have a nearly uniform distribution between zero and $4e^2/h$. For the robustness $R$, we find that most of the samples have vanishing robustness, and only a few samples have slightly larger robustness up to $R\sim 0.02$. 

However, when we couple the two quantum dots ($t=1$), as shown in Fig.~\ref{fig:3}(d) {where we show all the samples with $Q>-0.95$}, we find that the dome shape of the joint distribution becomes similar to the single quantum dot with two leads in Fig.~\ref{fig:3}(b), where the large and small zero-bias conductances rarely occur, and most of the zero-bias conductances appear near $2e^2/h$. The large zero-bias conductance can also possess strong robustness up to $R\sim 0.04$, and the robustness generally decreases as one goes away from $G\sim 2e^2/h$. However, one salient difference is that the concentration of the samples with zero-bias conductances near $2e^2/h$ is much steeper in the $G$ axis and has a longer tail in the $R$ axis up to $R\sim 0.05$ than that in Fig.~\ref{fig:3}(b). In addition, the extent of the dome shape on the $G$ axis is also wider. Again, this points to the unfortunate fact that disorder-induced trivial ZBCPs with the ``quantized'' strength $\sim 2e^2/h$ appear often to have significant ``robustness'', thus leading to false positives of them being considered to be ``topological'' MZMs.

Besides the chain of quantum dots with $L=2$, we also provide two other configurations with more quantum dots in Fig.~\ref{fig:3}(e) for $t=0.5$, and in Fig.~\ref{fig:3}(f) for $t=1$ (more results are shown in Appendix~\ref{app:A}). 
We first confirm the generic behavior of the joint distribution for $R$ and $G$, which manifests a dome-shaped probability density function as before. However, the support on the $G$ axis near $G\sim2e^2/h$ is also different from the single quantum dot in Fig.~\ref{fig:3}(b), which shows the narrowest support ($G\in[0.5e^2/h,3e^2/h]$), while the chain of quantum dots in Fig.~\ref{fig:3}(f) shows a much larger support (whose marginal distribution for $G$ almost becomes a uniform distribution) similar to the decoupled quantum dots in Fig.~\ref{fig:3}(c). 

This trend of the dome shape can be understood in terms of the effective length of the system, i.e., the ratio of the absolute total length $L$ to its {coherence length}. {The coherence length $\xi$ is defined in terms of the inverse of the Lyapunov exponent, where the Lyapunov exponent is $\xi^{-1}=\frac{1}{L}\log(\abs{\lambda})$.  Here $L$ is the total length and $\lambda$ is the largest eigenvalue of the transmission matrix $t t^\dagger$~\cite{pekerten2017disorderinduced} [where $t$ is the off-diagonal block of the scatter matrix in Eq.~\eqref{eq:mw}]. 
Here we want to understand the trend in the coherence length as $t$ is varied. This will allow us to understand the trend in the effective length $L/\xi$ as $t$ is changed. The coherence length $\xi$ of the disordered Kitaev chain is known to decrease as $t$ is increased~\cite{hegde2016majorana}.}
For example, Fig.~\ref{fig:3}(e) with $t=0.5$ and $L=5$ has a larger effective length than Fig.~\ref{fig:3}(f) with $t=1$ and $L=5$ due to a smaller {coherence length}. Similarly, Fig.~\ref{fig:3}(f) with $t=1$ and $L=5$ has a larger effective length than Fig.~\ref{fig:3}(d) with $t=1$ and $L=2$ because they have comparable {coherence lengths} but the absolute total length $L=5$ is longer than $L=2$. 
For the single quantum dot in Fig.~\ref{fig:3}(b), this can be viewed as a system with an extremely short effective length.
However, for the decoupled quantum dots in Fig.~\ref{fig:3}(c), this can be viewed as a system of an infinite wire because the vanishing {coherence length} causes the divergence of the effective length.
Thus, we conclude that a shorter effective length will generate more trivial ZBCPs showing $2e^2/h$ with stronger robustness. For the longer effective length, this trend is less prominent: the trivial ZBCPs have an almost uniform distribution of the zero-bias conductance.

\subsection{Normalized mutual information}

\begin{figure}[ht]
    \centering
    \includegraphics[width=3.4in]{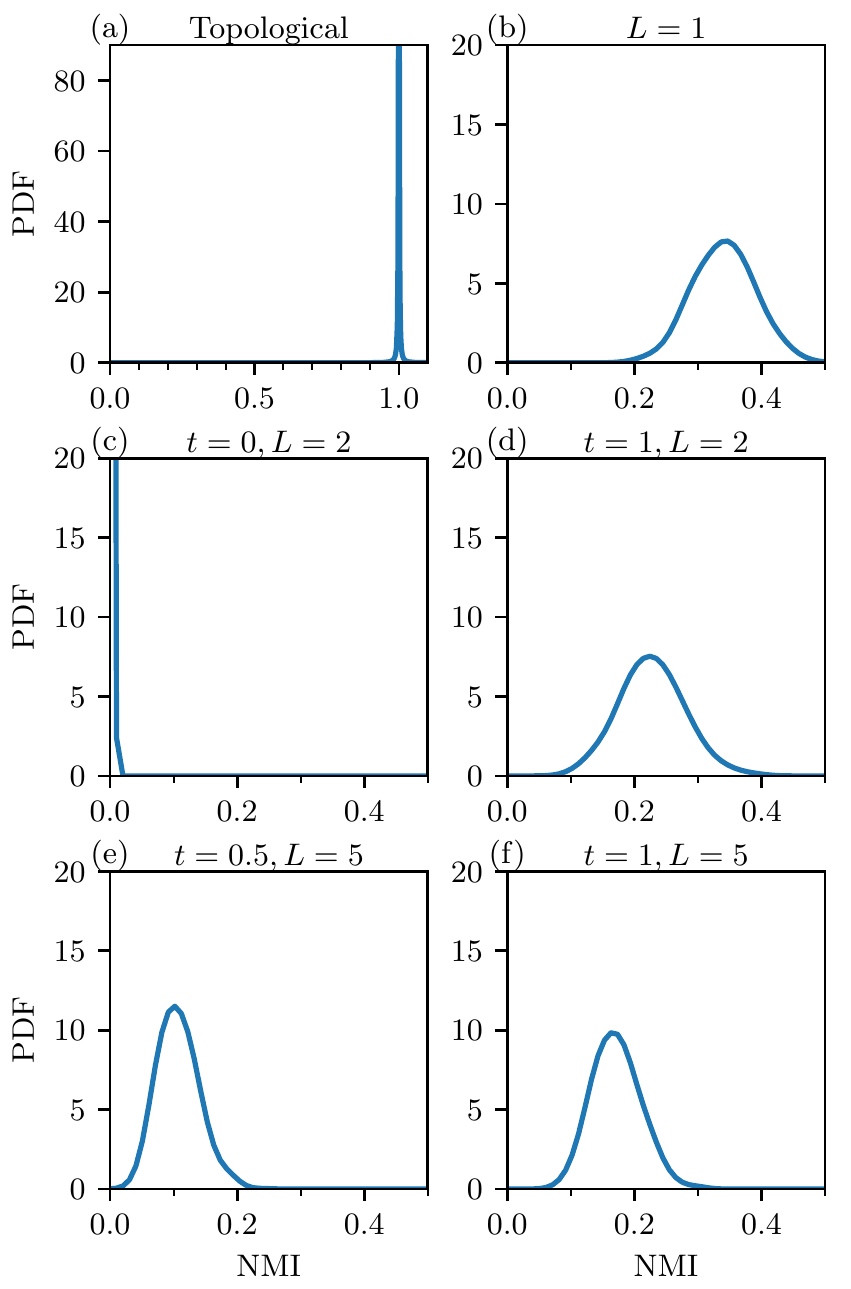}
    \caption{The probability density function of the normalized mutual information for (a) a topological case (as a reference), (b) a single quantum dot with two leads, (c) two quantum dots with no coupling, (d) two quantum dots with $t=1$, (e) five quantum dots with $t=0.5$, and (d) five quantum dots with $t=1$. The ensemble size for each case is 1000.}
    \label{fig:4}
\end{figure}

Besides the joint distribution between the robustness $R$ and the zero-bias conductance $G$, we also study the correlation between the zero-bias conductances from both leads.
Here, we first generate a set of $(H_1,H_2,H_3)$ and sweep $\alpha_1$ and $\alpha_2$ to obtain two ``phase diagrams'' of the zero-bias conductances from both leads, e.g., Figs.~\ref{fig:1}(e) and~\ref{fig:1}(f). 
We then separate trivial phases and topological phases according to the {topological visibility} $Q$,  
and individually calculate the NMI between the zero-bias conductances from both leads for trivial phases and topological phases (if any). This pair of NMI (for topological phases and trivial phases) is considered one sample associated with the set $(H_1,H_2,H_3)$ in the entire ensemble (for topological phases and trivial phases). 
Next, we change to another set of $(H_1,H_2,H_3)$, and repeat the same procedure to obtain a second pair of NMI as another sample. 
We randomize the ensemble by choosing 1000 different sets of $(H_1,H_2,H_3)$, and plot their PDF as a function of NMI using KDE as shown in Fig.~\ref{fig:4}.

We first present the result of a topological case as a reference in Fig.~\ref{fig:4}(a), where we find that the NMI is almost a Dirac $\delta$ function peaking at 1 as expected. Because the topological phase is bound to manifest a {``quantized''} conductance of $2e^2/h$, all the conductance pairs from both leads have the same value of $2e^2/h$, which further results in both zeros of the mutual information and joint entropy. Since the NMI is a quantity that measures the ratio of the mutual information to the total information, it is reasonable to define $0/0=1$, the ``idea'' value, for the NMI in the topological case.

With the reference to the topological case, we then present the result of the single quantum dot in Fig.~\ref{fig:4}(b), and find that the distribution of NMI peaks at 0.3. Since the two leads probe the same quantum dot, the correlation between the two leads is strong as expected for a very short system. 

Furthermore, we also present the results of the chain of quantum dots in Figs.~\ref{fig:4}(c)-\ref{fig:4}(f). Figure~\ref{fig:4}(c) shows the ensemble from decoupled double quantum dots, where the NMI is trivially zero because they are just two independent quantum dots manifesting completely irrelevant conductances. However, when we turn on the couplings between quantum dots in the chain, as shown in Figs.~\ref{fig:4}(d)-\ref{fig:4}(f), we find that the mode of the PDF of NMI decreases as the effective length increases. This indicates that the end-to-end correlation of the trivial ZBCPs vanishes as the effective length approaches infinity. Unfortunately, for wires of finite lengths, even the trivial ZBCPs would manifest some end-to-end correlations depending on the details.

\begin{figure}[ht]
    \centering
    \includegraphics[width=3.4in]{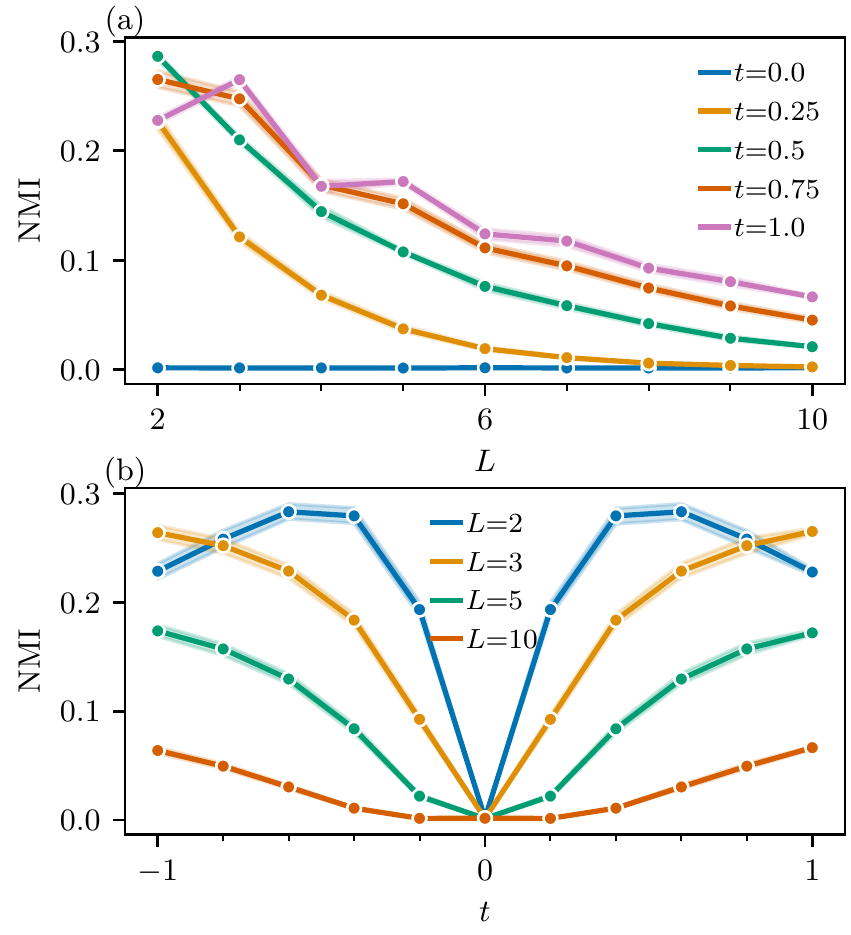}
    \caption{The mean normalized mutual information (a) at a fixed length $L$ for different hoppings $t$, and (b) at a fixed hopping $t$ for different lengths $L$. The shaded regions indicate the 95\% confidence interval in an ensemble of 1000.}
    \label{fig:5}
\end{figure}

Finally, we systematically study the NMI by varying the length $L$ between 2 and 10, and the hopping $t$ between $-1$ and $+1$ in Fig.~\ref{fig:5}. 
In Fig.~\ref{fig:5}(a), we plot the average NMI (solid dots) along with the 95\% confidence interval (shaded region) for the trivial phase as a function of $L$ for different hoppings $t$. The 95\% confidence interval is determined based on $t$-statistics of an ensemble of 1000.  
We find that the NMI remains zero at $t=0$ for all lengths $L$ because these are the cases with infinite effective lengths. 
However, for a finite $t$, the mean NMI and 95\% confidence interval decreases as $L$ increases because of the larger effective length. 

We also fix the number of quantum dots in the chain, and plot the mean NMI and 95\% confidence interval as a function of $t$ in Fig.~\ref{fig:5}(b). We find that the NMI is symmetric around the axis of $t=0$ because of the particle-hole symmetry. 
For a longer chain (e.g., $L=5$ and $10$), the average and variance of NMI increase as $\abs{t}$ increases. However, for a chain with fewer quantum dots, the average and variance of NMI increase until $\abs{t}$ reaches 0.4, beyond which they decrease. 
{To understand this nonmonotonic trend, we note that the NMI is related to the transmission from one end to the other end; i.e., if no transmission can happen, then the NMI is zero, while the NMI is large if given a large transmission.
The transmission can be further determined by the dot-dot hopping $t$, which can be more physically characterized by the thermal conductance. 
In general, the dot-dot thermal conductance increases with the increase of the dot-dot hopping $t$. [We present the mapping between the thermal conductance and the hopping $t$ in a chain of two quantum dots (QDs) in Appendix~\ref{app:C} along with the details of the methodology.]
However, the transmission between the two dots does not solely determine the total transmission because the bottleneck of the total transmission is determined by the smallest thermal conductance among the two lead-dot thermal conductances and dot-dot thermal conductance (which satisfies the series parallel resistor rules due to the conservation). Therefore, when $t$ is near zero, because the lead-dot conductance is much larger than the dot-dot conductance, we see an increase in total transmission contributed by the increase of $t$. However, the increase of total transmission saturates when the dot-dot conductance is of the same order as the lead-dot conductance, beyond which the bottleneck of the total transmission becomes the lead-dot conductance. In Appendix~\ref{app:C}, we have numerically verified that the crossover happens near $t=0.4$ at $L=2$.}

\section{discussion and conclusion}\label{sec:conclusion}

In this paper, we use the random matrix theory in a class D ensemble to simulate the transport properties of quantum dots and wires by analogy with the superconductor-semiconductor nanowire. 
Our work is complementary to the extensive existing theoretical literature in Majorana nanowires based on microscopic direct solutions of the BdG equation in the presence of Zeeman field and spin-orbit coupling. 
We consider two types of models: the single quantum dot with two leads, and the chain of quantum dots, which is generalized from the single quantum dot by explicitly adding the concept of length and hopping to the system. 

We first calculate the conductance using the Mahaux-Weidenm\"uller formula, treating the two leads as one large lead to calculate the corresponding $S$ matrix. By introducing the tuning parameter $\vec{\alpha}$, we can manipulate the Hamiltonian in a similar way to the fine-tuning in experiments. We, therefore, visualize the zero-bias conductances as a function of $\alpha_1$ and $\alpha_2$ in the 2D false-color ``phase diagrams'', and find that both the single quantum dot and chain of quantum dots can manifest similar ``phase diagrams'' of zero-bias conductances.

We then determine the existence of ZBCPs from both leads by identifying purely imaginary eigenvalues in a non-Hermitian effective Hamiltonian, and define the robustness based on the center-to-boundary distance of a ZBCP. From a large random ensemble, we separate the samples of topological phases and trivial phases, and find that topological phases always robustly carry the {``quantized''} conductance of $2e^2/h$ while trivial phases manifest a strong correlation between the robustness and zero-bias conductance in the plateaulike region where ZBCPs exist: the trivial ZBCPs with stronger robustness are prone to manifest larger zero-bias conductances near $2e^2/h$ while the trivial ZBCPs with weak robustness usually manifest vanishing zero-bias conductances. This correlation becomes even more significant in shorter chains. On the contrary, in the long chain limit, the distribution of the zero-bias conductances conditioned on the robustness becomes featureless such that all samples are almost uniformly distributed regardless of the robustness. The long chains also tend to host ZBCPs with weaker robustness than the short chains.
One important implication of these results is that, in short experimental wires, there would be basically no difference between trivial and topological, and experiments should focus on long wires, where an underlying topology is meaningful.

Finally, we study the end-to-end correlation of the zero-bias conductances between the two leads by calculating their NMI. For topological phases, the NMI has to be exactly 1 because both ends can only give identical zero-bias conductances of $2e^2/h$.
However, for the trivial phase, we find that the average NMI and the 95\% confidence interval of NMI both decrease as the effective length increases.  For the single quantum dot, the NMI shows the largest value up to 0.3 whereas it becomes vanishing for the chain of quantum dots with effectively infinite lengths. Thus, in longer wires, a meaningful distinction can be made between trivial and topological MZMs.

Our paper is motivated by the fact that experimental data, even when they arise from trivial class D disorder physics, can manifest some degrees of the robustness of ZBCPs, quantized zero-bias conductances, and even end-to-end correlation of the zero-bias conductances from both leads, which can sometimes resemble the signatures created by the topological MZMs. However, we show that these hallmarks of the topological MZMs should also be treated with caution if observed in experiments as they can be misrepresented by trivial ZBCPs to some degree as well. Therefore, the mere observations of these hallmarks to some extent do not decisively indicate the existence of topological MZMs. In real experiments, one has to confirm that the evidence of MZMs is statistically more significant than the signatures induced by trivial ZBCPs before claiming the observation of the topological MZMs.
However, if all features above can appear simultaneously in all the samples, it will be a strong indication of the topological MZMs. One decisive conclusion is that experiments should report extensive statistical analyses of ZBCPs instead of fine-tuning cherry-picked ZBCPs with the expected $2e^2/h$ conductance.

Because our work relies on the random matrix Hamiltonian in a class D ensemble without referring to any explicit experimentally relevant model, one advantage is that the results should be generic for any platform searching for topological MZMs with disorder and finite-size effect, not just to the SC-SM nanowire platforms. However, being generic also means that our work does not give an explicit threshold that can be directly mapped into specific experimental platforms to decisively determine the topological versus trivial ZBCPs. Nevertheless, we provide a detailed generic protocol that provides guidelines for such a statistical test, which can increase the likelihood of arriving at a correct conclusion. 

Although we believe that much of the existing experimental literature reporting on zero-bias conductance peaks in the local tunneling spectroscopy of Majorana nanowires is qualitatively consistent with our random matrix theory analysis presented in the current work (because of the high degree of unintentional quenched disorder present in the nanowires), it may be interesting to ask if there is a direct way of quantitatively testing our predictions. One straightforward experimental implementation of our idea would be to produce a one-dimensional (1D) system using randomly placed InAs or InSb quantum dots, simulating a random Kitaev chain (except made of coupled semiconductor quantum dots here) in the presence of an Al film to provide proximity superconductivity and an applied magnetic field in order to produce spin polarization. Basically, the system is the usual semiconductor nanowire, but the nanowire is made of randomly placed quantum dots.  Such a system directly emulates a class D random system, and should be well described by our theory, and the experimentally studied statistics of tunneling zero-bias conductance peaks in this random ``Kitaev chain'' can be compared directly with our theory.  
In fact, such a system with a 1D quantum dot lattice was proposed more than a decade ago~\cite{sau2012realizing}, 
and recently such an artificial semiconductor dot-based ``Kitaev chain'' has been realized~\cite{dvir2022realization} experimentally.  
Therefore, fabricating a random 1D chain of quantum dots should be feasible with the existing fabrication technique, leading directly to the verification of our random matrix theory for the tunnel conductance zero-bias peak class D statistics.

\section*{Acknowledgements}
This work is supported by the Laboratory for Physical Sciences. We thank the University of Maryland High-Performance Computing Cluster (HPCC) for providing computational resources.

\bibliography{Paper_RM2}
\appendix
\onecolumngrid
\section{More examples of the joint distribution for $R$ and $G$}\label{app:A}
\begin{figure}[ht]
    \centering
    \includegraphics[width=6.8in]{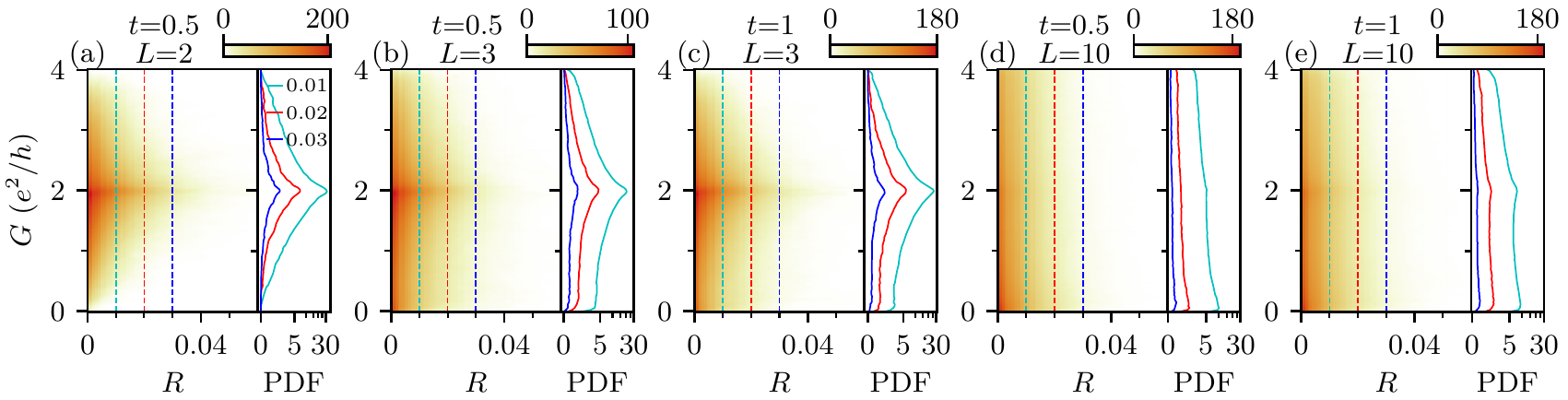}
    \caption{The joint distribution for the robustness $R$ and the zero-bias conductance $G$ for (a) two quantum dots with $t=0.5$, (b) three quantum dots with $t=0.5$, (c) three quantum dots with $t=1$, (d) ten quantum dots with $t=0.5$, and (e) ten quantum dots with $t=1$. The solid lines on the right of each panel show the conditional probability density function at a fixed robustness $R=0.01$ (cyan), $R=0.02$ (red), and $R=0.03$ (blue). The ensemble size for each case is 1000.}
    \label{fig:AppA}
\end{figure}

In this section, we present more results of the PDF of joint distribution as a function of the robustness $R$ and the zero-bias conductance $G$ for different configurations of $t$ and $L$ in Fig.~\ref{fig:AppA}.

\section{More examples of the normalized mutual information}\label{app:B}
\begin{figure}[ht]
    \centering
    \includegraphics[width=6.8in]{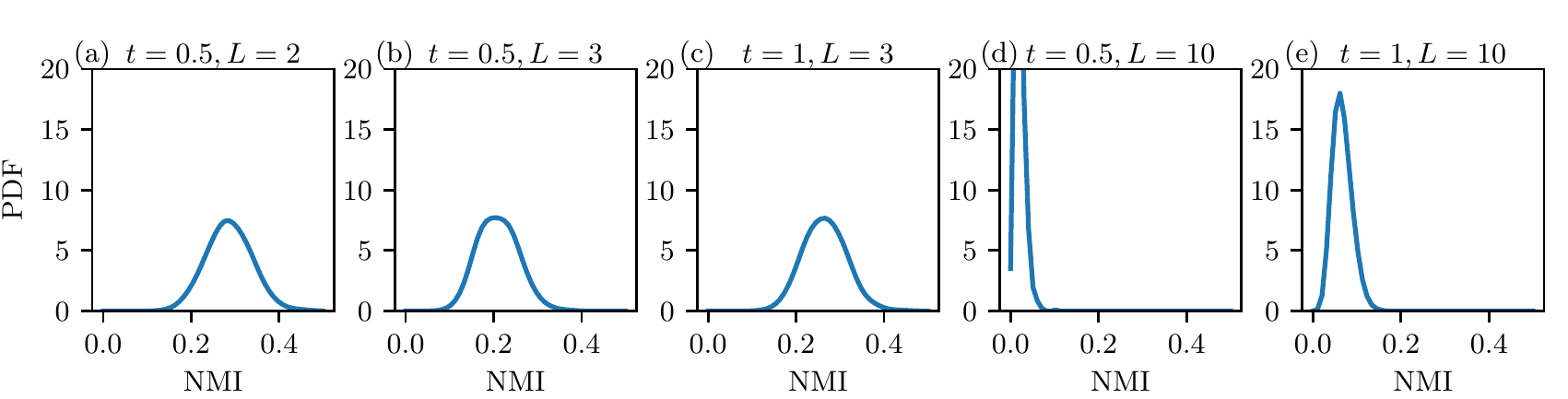}
    \caption{The probability density function of the normalized mutual information obtained for (a) two quantum dots with $t=0.5$, (b) three quantum dots with $t=0.5$, (c) three quantum dots with $t=1$, (d) ten quantum dots with $t=0.5$, and (e) ten quantum dots with $t=1$. The ensemble size for each case is 1000.}
    \label{fig:AppB}
\end{figure}

In this section, we present more results of the PDF of the normalized mutual information for different configurations of $t$ and $L$ in Fig.~\ref{fig:AppB}.

\section{The comparison between the lead-dot thermal conductance and dot-dot thermal conductance}\label{app:C}

We compute the thermal conductance by counting the transmission of all types, i.e.,
\begin{equation}
    \kappa=T_{ee}+T_{eh},
\end{equation}
where $T_{ee}=\tr(t_{ee}t_{ee}^\dagger)$ and $T_{eh}=\tr(t_{eh}t_{eh}^\dagger)$. Here $t$ is the transmission matrix, i.e., the off-diagonal block of the scattering matrix $S(E=0)$ in Eq.~\eqref{eq:mw} (the direction does not matter here as they are always the same). 

To extract the dot-dot thermal conductance from the total transmission (thermal conductance), we require the lead-dot thermal conductance to be as large as possible such that it will not be the bottleneck of the total transmission. Namely, the total thermal conductance is (assuming an Ohmic limit)
\begin{equation}\label{eq:series}
    \kappa^{-1}\approx\kappa_{\text{L-dot}}^{-1}+\kappa_{\text{R-dot}}^{-1}+ \kappa_{\text{dot-dot}}^{-1},
\end{equation}
where $\kappa_{\text{L-dot}}=\kappa_{\text{R-dot}}$ are the (left and right) lead-dot thermal conductance, and $\kappa_{\text{dot-dot}}$ is the dot-dot thermal conductance. 
It is only when $\kappa_{\text{L-dot}}\rightarrow\infty$ that we have $\kappa_{\text{dot-dot}}\approx \kappa$.

\begin{figure}[ht]
    \centering
    \includegraphics[width=6.8in]{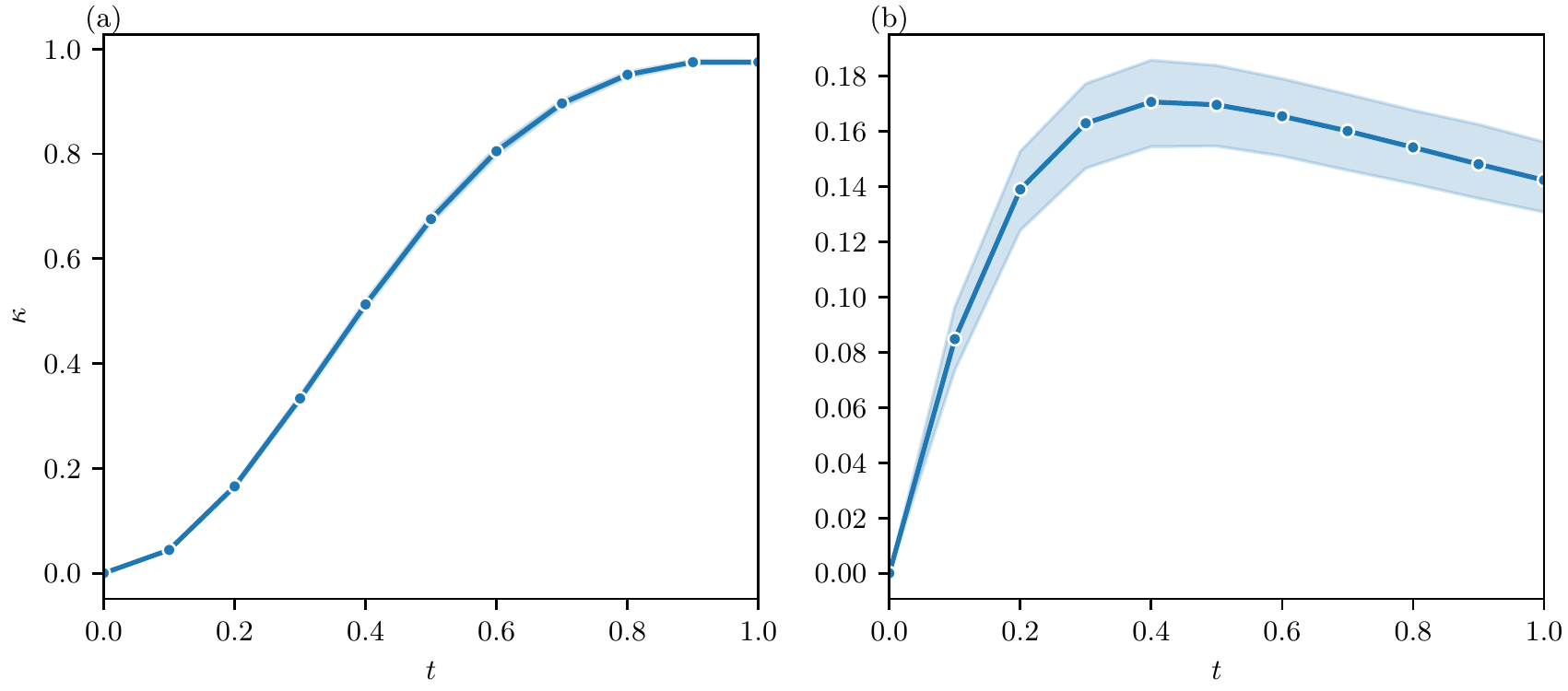}
    \caption{(a) The total thermal conductance (which is also the dot-dot thermal conductance) as a function of the dot-dot hopping $t$ for an $L=2$ chain with $\Gamma_n=1$ and $N=80$. The shaded area is the standard error in an ensemble of 100000. (b) The total thermal conductance as a function of the dot-dot hopping $t$ for an $L=2$ chain with $\Gamma_n=0.1$ and $N=4$. The lead-dot conductance is estimated to be approximately 0.3 taking into account the two junctions.}
    \label{fig:AppC}
\end{figure}

Therefore, we increase the tunneling probability $\Gamma_n$ to 1 and the number of channels in the lead $N$ to 80 to reach the large limit of $\kappa_{\text{L-dot}}\rightarrow\infty$, and present the result of the total transmission in Fig.~\ref{fig:AppC}(a), which can also be thought of as the dot-dot transmission. {The bounded value of the conductance at near $\kappa\sim 1$ is due to the finite $M=80$. In principle, one can achieve higher conductance by setting a larger $M$ and $N$.}

Once we estimate the dot-dot thermal conductance, we can estimate the lead-dot thermal conductance at a smaller $\Gamma_n$. We first numerically calculate the total transmission for the same configurations but a smaller $\Gamma_n=0.1$ (used in the main text to produce all the figures) as shown in Fig.~\ref{fig:AppC}(b).
We find that the total transmission is around 0.14 to 0.18 at a large $t$. Assuming we can ignore the dot-dot coupling in this range, the lead-dot conductance in one junction is twice the total conductance, which is between 0.28 and 0.36, which is close to the dot-dot conductance ($\kappa\sim0.5$) at $t=0.4$, indicating the crossover between a chain of two QDs and an effective single large quantum dot.

\end{document}